# Tracking Target Signal Strengths
# on a Grid using Sparsity

Shahrokh Farahmand, Georgios B. Giannakis, Geert Leus, and Zhi Tian

### Abstract

Multi-target tracking is mainly challenged by the nonlinearity present in the measurement equation, and the difficulty in fast and accurate data association. To overcome these challenges, the present paper introduces a grid-based model in which the state captures target signal strengths on a known spatial grid (TSSG). This model leads to *linear* state and measurement equations, which bypass data association and can afford state estimation via sparsity-aware Kalman filtering (KF). Leveraging the grid-induced sparsity of the novel model, two types of sparsity-cognizant TSSG-KF trackers are developed: one effects sparsity through $\ell_1$-norm regularization, and the other invokes sparsity as an extra measurement. Iterative extended KF and Gauss-Newton algorithms are developed for reduced-complexity tracking, along with accurate error covariance updates for assessing performance of the resultant sparsity-aware state estimators. Based on TSSG state estimates, more informative target position and track estimates can be obtained in a follow-up step, ensuring that track association and position estimation errors do not propagate back into TSSG state estimates. The novel TSSG trackers do not require knowing the number of targets or their signal strengths, and exhibit considerably lower complexity than the benchmark hidden Markov model filter, especially for a large number of targets. Numerical simulations demonstrate that sparsity-cognizant trackers enjoy improved root mean-square error performance at reduced complexity when compared to their sparsity-agnostic counterparts.

**EDICS:** SSP-TRAC.

S. Farahmand and G. B. Giannakis are with the Dept. of Electrical and Computer Engineering, University of Minnesota, Minneapolis, MN, 55455 USA. e-mails: {shahrokh,georgios}@umn.edu; G. Leus is with the Faculty of Electrical Engineering-Mathematics-Computer Science, Delft University of Technology, Delft, Netherlands, email: g.j.t.leus@tudelft.nl; Z. Tian is with the Dept. of Electrical and Computer Engineering, Michigan Tech University, Houghton, MI, USA. email: ztian@mtu.edu.

The first two authors were supported by the NSF grants CCF-0830480, CCF-1016605, ECCS-0824007, and ECCS-1002180. Geert Leus was supported in part by NWO-STW under the VICI program (project 10382). Zhi Tian was partly supported by the NSF grant ECS-0925881. Part of the results in this paper will appear in *Proc. of Intl. Conf. on Information Fusion*, Chicago, IL, July 2011.

# I. INTRODUCTION

Target tracking research and development are of major importance and continuously expanding interest to a gamut of traditional and emerging applications, which include radar- and sonar-based systems, surveillance and habitat monitoring using distributed wireless sensors, collision avoidance modules envisioned for modern transportation systems, and mobile robot localization and navigation in static and dynamically changing environments, to name a few; see e.g., [4], [9], and references therein.

At the core of long-standing research issues even for single-target tracking applications is the *nonlinear* dependence of the measurements on the desired state estimates, which challenges the performance of linearized Kalman filter (KF) trackers, including the extended (E)KF, the unscented (U)KF, and their iterative variants [4], [9]. This has motivated the development of particle filters (PFs), which can cope with nonlinearities but tend to incur prohibitively high *complexity* in many critical applications. For multi-target tracking, *data association* has been another formidable challenge, especially when the ambient environment is cluttered, and the sensors deployed are unreliable. This challenge amounts to determining the target associated with each measurement, where the noisy measurements typically reflect the candidate target locations acquired through signal detection in gated validation regions; see e.g., [3], [9]. Once data association is established, targets can be tracked separately using the associated measurements, in conjunction with track fusion for improved accuracy.

The present paper investigates the multi-target tracking problem whereby the available measurements comprise the superposition of received target signal strengths of all targets in the sensor field of view. Sensors collecting these measurements are not necessarily radars or high-cost receivers, but can be general-purpose radio units employing simple energy detectors. The measurements are nonlinearly related to target locations but no data association issues arise, because conventional range-gate operations have not yet been employed to detect, separate, and localize the targets of interest [3]. To cope with the nonlinearity issue, this paper introduces a grid-based dynamical state-space model in which the state describes signal strengths of targets traversing a preselected spatial grid (TSSG) of the tracking field. Because the locations of grid points are preset and known, both the measurement and state equations become linear. Further, data association is avoided by dynamically tracking the TSSG values rather than directly producing the target tracks. Based on TSSG tracking however, data association and track trajectory estimation can be performed as a follow-up step, whereby track association and estimation errors do not propagate back to the TSSG tracker.

Similar ideas on bypassing data association at the price of tracking "less informative" estimates have been exploited in recent multi-target tracking schemes, such as the probability hypothesis density (PHD) filter [22], [30] and the Bayesian occupancy filter (BOF) [14]. The PHD filter tracks the so-termed target intensity, while the BOF tracks the probability of a grid point being occupied by any target. A main advantage of the *grid-based* TSSG tracker here is that state estimation becomes possible via KF applied to *linear* state and measurement models, at considerably reduced computational burden relative to the complexity incurred by the PHD and BOF. Further, the TSSG tracker is novel in exploiting the *sparsity* present in the grid-based state vector, which allows one to leverage efficient solvers of (weighted) least-squares (LS) minimization problems regularized by the $\ell_1$-norm of the desired state estimate.

Sparsity-aware estimators have been studied for variable selection in *static* linear regression problems, and have recently gained popularity in signal processing and various other fields in the context of compressive sampling (CS); see e.g., [5], [11], [21]. However, few results pertain to the *dynamic* scenario encountered with target tracking. When measurements arrive sequentially in time, a sparsity-aware recursive least-squares scheme was reported in [1], but its tracking capability is confined only to slow model variations; see also [2] for a sparsity-cognizant smoothing scheme which nevertheless does not lend itself to filtering; as well as [29], where a so-called KF-CS-residual scheme is reported for tracking slowly varying sparsity patterns. Different from existing alternatives, the present work develops sparsity-aware trackers along with their error covariances, without requiring knowledge on the number of (possibly fast-moving) targets or their signal strengths.

The rest of the paper is organized as follows. Section II develops the novel grid-based sparse model, for which a sparsity-agnostic KF tracker is introduced in Section III. Two sparsity-cognizant trackers are presented in Sections IV and V. Target position estimation and track formation is detailed in Section VI. Numerical results are presented in Section VII, followed by concluding remarks in Section VIII.

## II. GRID-BASED STATE SPACE MODEL

Consider the problem of tracking $M$ moving targets using $N$ active (e.g., radar) or passive (e.g., acoustic) sensors deployed to provide situational awareness over a geographical area. Targets emit power either because they passively reflect the energy of other transmitters such as radar, or, because they are active sources such as cell-phones or transmitters mounted on smart cars. Associated with each target, say the $m$th one, is its position vector $\mathbf{p}_k^{(m)}$ per time $k$, and the signal of strength $s^{(m)}$ that the target reflects or emits. Sensor $n$ measures the superposition of received target signal strengths,

$$y_{n,k} = \sum_{m=1}^{M} h\left(d_k^{m \to n}\right) s^{(m)} + \nu_{n,k}, \quad n = 1, \ldots, N, \quad k = 1, 2, \ldots \tag{1}$$

where $h(\cdot)$ denotes the distance-dependent propagation function; $d_k^{m \to n} := \|\mathbf{p}_k^{(m)} - \mathbf{q}_n\|_2$ is the distance between the known position $\mathbf{q}_n$ of sensor $n$ and the unknown position vector $\mathbf{p}_k^{(m)}$ of target $m$; and $\nu_{n,k}$ is zero-mean Gaussian noise at sensor $n$. Function $h(\cdot)$ satisfies $h(0) = 1$, is non-negative, decreasing, and is either assumed known from the physics of propagation or acquired through training [21].

At each time $k$, a centralized processor has available the measurement vector $\mathbf{y}_k := [y_{1,k}, \ldots, y_{N,k}]^T$, based on which the target positions $\{\mathbf{p}_k^{(m)}\}_{m=1}^M$ are to be tracked. Note that the measurement model (1) differs from the one typically considered in radar applications, where a measurement either comes from a single target or a clutter, usually in the form of position information obtained from range gate operations [3]. Each measurement in (1) comes from a sensor, and comprises the superposition of received signal strengths emitted by or reflected from all targets in the sensor field of view. This model considers the localization and tracking problems jointly, and avoids the measurement-target association issue.

One major challenge in tracking and localization problems is that the measurements in (1) are nonlinear functions of the wanted target position vectors. A neat approach to arrive at a linear measurement model is to adopt a set of $G$ (possibly regularly spaced) grid points at known positions $\{\mathbf{g}_i\}_{i=1}^G$, where target(s) could be potentially located; see also e.g., [14], [11], and [5]. Using a sufficiently dense grid, it is possible to capture the target locations at a prescribed spatial resolution using a $G \times 1$ vector $\mathbf{x}_k$ having most entries equal to zero except for the $\{i_k^{(m)}\}_{m=1}^M$ entries given by $\{x_k^{(i_k^{(m)})}\}_{m=1}^M$, which represent the target signal strengths at time $k$ if and only if the $m$-th target is located at the $i_k^{(m)}$-th grid point, that is $\mathbf{p}_k^{(m)} = \mathbf{g}_{i_k^{(m)}}$. Note that if target $m$ is located exactly on a grid point $i_k^{(m)}$, then $x_k^{(i_k^{(m)})} \equiv s^{(m)} \neq 0$ will be the only nonzero entry of $\mathbf{x}_k$ corresponding to this target. However, to account for target presence off the preselected grid points, it will be allowed for the unknown target signal strength $s^{(m)}$ to "spill over" grid points around $i_k^{(m)}$ and thus render nonzero a few neighboring entries of $\mathbf{x}_k$. Let $\mathcal{G}_k^{(m)}$ denote the spill-over region on the grid corresponding to target $m$ at time $k$, such that $x_k^{(i)} \neq 0$ is associated with $s^{(m)}$, $\forall i \in \mathcal{G}_k^{(m)}$. The following assumption on this target occupancy model is imposed:

**(as1)** *Each grid point $i$ can be occupied by at most one target $m$ at any given time $k$.*

This assumption can be easily satisfied in practice by selecting a sufficiently dense grid [14], [15]. Under as1), each grid point $i$ is associated with a unique target index $m_k^{(i)}$ at time $k$; that is, $i \in \mathcal{G}_k^{(m_k^{(i)})}$, where $m_k^{(i)} \in [1, M]$ if it is occupied by one of the $M$ targets; or, $m_k^{(i)} = 0$ if it is not occupied, meaning it is associated with a dummy target $m = 0$ with strength $s^{(0)} \equiv 0$. Apparently, $\{\mathcal{G}_k^{(m)}\}_{m=0}^M$ are mutually exclusive across $m$ and their union spans the entire grid in the sense $\cup_{m=0}^M \mathcal{G}_k^{(m)} = \cup_{i=1}^G i$, which leads to a measurement equation [cf. (1)]

$$y_{n,k} = \sum_{m=0}^{M} \sum_{i \in \mathcal{G}_k^{(m)}} h\left(d^{(i \to n)}\right) x_k^{(i)} + v_{n,k} = \mathbf{h}_n^T \mathbf{x}_k + v_{n,k}. \tag{2}$$

Here $\mathbf{h}_n^T := [h(d^{1 \to n}), h(d^{2 \to n}), \dots, h(d^{G \to n})]$; $d^{i \to n} := \|\mathbf{q}_n - \mathbf{g}_i\|_2$ now denotes the known time-invariant distance between the $n$th sensor and the $i$th grid point; and the noise $v_{n,k}$ replacing $\nu_{n,k}$ in (1) captures the unmodeled dynamics in the aforementioned spill-over effect. Notwithstanding, thanks to the grid-based model, the measurements in (2) have become linear functions of the unknown $\mathbf{x}_k$ whose nonzero entries reveal the grid points where target signal strengths are present at time $k$.

The next step is to model the evolution of $\mathbf{x}_k$ in time as the targets move across the grid. Regarding their movement pattern, targets obey the following assumption:

**(as2)** *All targets move according to identical transition probabilities* $\{f_k^{(ji)}\}_{i,j=1}^G$, *where* $f_k^{(ji)} := p(x_k^{(j)} \neq 0 | x_{k-1}^{(i)} \neq 0; j \in \mathcal{G}_k^{(m)}, i \in \mathcal{G}_{k-1}^{(m)})$, $m = 1, \dots, M$.

In words, the homogeneity of targets under as2) refers to the probability that a target $m$ moves from grid point $i$ at time $k-1$ to point $j$ at time $k$.

Consider now expressing each entry of $\mathbf{x}_k$ as $x_k^{(j)} = s_k^{(j)} \cdot p(x_k^{(j)} \neq 0)$, where $s_k^{(j)} = s^{(m_k^{(j)})} \in \{s^{(0)}, s^{(1)}, \dots, s^{(M)}\}$ denotes a nonnegative proportionality constant, and $p(x_k^{(j)} \neq 0)$ stands for the probability of a target to be present on grid point $j$ at time $k$. Essentially, each $x_k^{(j)}$ is associated with only one of the $(M+1)$ targets (including the dummy target $m = 0$ indexed by $m_k^{(j)}$, and $s_k^{(j)}$ is a proportionality constant in the sense that it takes on $(M+1)$ possible values $s^{(m)} = \sum_{j \in \mathcal{G}_k^{(m)}} x_k^{(j)}$, for $m = 0, 1, \dots M$.

Under as1) and as2), it is shown in the Appendix that the state obeys the following recursion

$$x_k^{(j)} = \sum_{i=1}^{G} f_k^{(ji)} x_{k-1}^{(i)}, \quad \forall j \in [1, G]. \tag{3}$$

Concatenating (3) for $j = 1, \dots, G$, and (2) for $n = 1, \dots, N$, one arrives at the *grid-based* model

$$\mathbf{x}_k = \mathbf{F}_k \mathbf{x}_{k-1} + \mathbf{w}_k \tag{4a}$$

$$\mathbf{y}_k = \mathbf{H}\mathbf{x}_k + \mathbf{v}_k \tag{4b}$$

where the $G \times G$ state transition matrix $\mathbf{F}_k$ has its $(i, j)$-th entry given by $f_k^{(ji)}$; the measurement matrix is defined as $\mathbf{H} := [\mathbf{h}_1, \dots, \mathbf{h}_n]^T$; likewise for the measurement noise vector $\mathbf{v}_k := [v_{1,k}, \dots, v_{N,k}]$; and $\mathbf{w}_k$ is a zero-mean process noise vector with a positive-definite covariance matrix $\mathbf{Q}_k$ added to account for both as1) and the natural non-negativity constraints on $\mathbf{x}_k$ whose entries represent target signal strengths (magnitudes or power).

A distinct feature of model (4) is that the unknown $\mathbf{x}_k$ is *sparse* $\forall k$, since only few of its $G$ entries are nonzero (in fact exactly $M$ nonzero entries if all the $M$ targets are located on grid points). Although (3) describes the linear evolution of each $\mathbf{x}_k$ entry under as1), using these recursions alone does not guarantee that the predicted or estimated $\mathbf{x}_k$ adheres to as1). Indeed, starting with a target at an arbitrary entry in $\mathbf{x}_0 \neq \mathbf{0}$ and running (3) up to a large enough $k$, the signal strength of this target will "spill-over" to all entries of $\mathbf{x}_k$, and will possibly overlap with other targets present. Such a state transition pattern is expected, because uncertainty of any dynamically evolving state grows over time if no corrections are made based on real-time measurements. Therefore, $\mathbf{x}_k$ predictions based on (4a) will be non-sparse, but the true state vector $\mathbf{x}_k$ at any time $k$ is sparse with only a few nonzero entries around the target locations. Posterior to processing the measurements, filtered and predicted renditions of $\mathbf{x}_k$ will remain sparse as well. The noise term $\mathbf{w}_k$ reflects the uncertainty in the state transition model under as1).

This sparsity attribute will prove to be instrumental for enhancing tracking performance. Also, it is worth noting that the state transition matrix $\mathbf{F}_k$ reflects the transition behavior of target positions only, without revealing full information of the target movement model that may be dependent on velocity or other factors as well. In fact, $\mathbf{F}_k$ is derived from the target movement model but does not fully reveal it, which differs from most existing track state models.

Given $\mathbf{y}_{1:k} := \{\mathbf{y}_1, \ldots, \mathbf{y}_k\}$, the goal of this paper is to track $\mathbf{x}_k$ using a *sparsity-aware* Kalman filter (KF). Since $\mathbf{x}_k$ represents the target signal strength on the grid (TSSG), the KF-like algorithms proposed in Sections III and IV will be referred to as TSSG–KF trackers, while the iterated extended Kalman filter (IEKF) algorithms of Section V will be referred to as TSSG–IEKF trackers. Having available $\hat{x}_k^{(j)}$ estimates, and recalling that $x_k^{(j)} = s^{(m_k^{(j)})} p(s_k^{(j)} \neq 0)$, one can estimate the constant $s^{(m)}$ capturing the signal strength of the $m$-th target at time $k$ as

$$\hat{s}_k^{(m)} = \sum_{j \in \mathcal{G}_k^{(m)}} \hat{x}_k^{(j)}, \quad \forall \ k \tag{5}$$

and the corresponding target position vector at time $k$ as

$$\hat{\mathbf{p}}_k^{(m)} = (1/\hat{s}_k^{(m)}) \sum_{j \in \mathcal{G}_k^{(m)}} \mathbf{g}_j \hat{x}_k^{(j)}, \quad m = 1, \ldots, M. \tag{6}$$

The following remark makes useful observations regarding the position estimate in (6).

**Remark 1.** A TSSG filter for tracking $\mathbf{x}_k$ avoids data association, because the TSSG-based state and measurement equations in (4) hold for any target-grid association $\{\mathcal{G}_k^{(m)}\}_m$, so long as as1) and as2) are satisfied. On the other hand, finding the target positions via (6) requires knowledge of $\{\mathcal{G}_k^{(m)}\}_m$, and hence calls for associating targets with TSSG entries. Solution to such an association problem will be provided in Section VI. Nonetheless, it is worth stressing that the association errors and resultant position estimation errors do not affect TSSG tracking that is independent of target position estimation, similar to the PHD and BOF in [22] and [14], respectively.

In addition to reduced complexity, an attractive feature of the present formulation relative to e.g., [14] is that even for finite $G$, there is no need to assume that targets are located on grid points since (6) allows for interpolating the target position vectors regardless, after knowing that grid point $j$ is associated with the target $m_k^{(j)}$ occupying it. The next remark is useful to further appreciate this point.

**Remark 2.** Given measurements $\mathbf{y}_{1:k}$, and supposing that the number of targets $M$ and their signal strengths $\{s^{(1)}, \ldots, s^{(M)}\}$ are known, the maximum a posteriori (MAP) and minimum mean-square error (MMSE) optimal trackers can be derived from a hidden Markov model (HMM) filter implementing the following recursions derived from Bayes' rule (cf. (34) and (35) in the Appendix)

$$p\left(x_k^{(j)} \neq 0 \,\Big|\, \mathbf{y}_{1:k-1}\right) = \sum_{i \in \mathcal{G}_{k-1}^{(m_k^{(j)})}} f_k^{(ji)} \, p\left(x_{k-1}^{(i)} \neq 0 \,\Big|\, \mathbf{y}_{1:k-1}\right)$$

$$p\left(x_k^{(j)} \neq 0 \,\Big|\, \mathbf{y}_{1:k}\right) = \frac{p(\mathbf{y}_k | x_k^{(j)} \neq 0; s^{(m_k^{(j)})}) p(x_k^{(j)} \neq 0 | \mathbf{y}_{1:k-1})}{\sum_{i \in \mathcal{G}_k^{(m_k^{(j)})}} p(\mathbf{y}_k | x_k^{(i)} \neq 0; s^{(m_k^{(i)})}) p(x_k^{(i)} \neq 0 | \mathbf{y}_{1:k-1})} \tag{7}$$

where $f_k^{(ji)}$ is the transition probability as in (3). These HMM recursions hinge on prior knowledge of the target-grid association $\{\mathcal{G}_k^{(m)}\}_{m=0}^{M}$, which need to be figured out among a total of $(M+1)^{G-M} G!/(G-M)!$ possible combinations. A large $G$ increases grid density and hence spatial resolution, at the expense of increasing complexity. In addition, $M$ and $\{s^{(m)}\}_{m=1}^{M}$ need to be known beforehand.

One additional remark is now in order.

**Remark 3.** Although $\mathbf{y}_k$ in (4b) comprises scalar measurements from $N$ geographically distributed sensors per time $k$, it is possible to form $\mathbf{y}_k$ with samples of the continuous-time signal received at a *single* (e.g., a radar or sonar) sensor by over-sampling at a rate faster than the rate $\mathbf{x}_k$ changes, so long as the state-space model (4) is guaranteed to be observable (and thus $\mathbf{x}_k$ is ensured to be identifiable).

## III. KF FOR TRACKING TSSG

If the non-negativity constraints for $\mathbf{x}_k$ were absent, the optimal state estimator for (4) in the MAP, MMSE, or least-squares (LS) error sense would be the clairvoyant linear KF. A pertinent state estimator is pursued here in the presence of non-negativity constraints. Suppose that the estimate $\hat{\mathbf{x}}_{k-1|k-1}$ and its error covariance matrix $\mathbf{P}_{k-1|k-1}$ are available from the previous time step. At time $k$, the KF state

predictor and its error covariance are obtained as

$$\hat{\mathbf{x}}_{k|k-1} = \mathbf{F}_k \hat{\mathbf{x}}_{k-1|k-1}$$
$$\mathbf{P}_{k|k-1} = \mathbf{F}_k \mathbf{P}_{k-1|k-1} \mathbf{F}_k^T + \mathbf{Q}_k. \tag{8}$$

For the KF corrector update, consider the LS formulation of the KF; see e.g., [27]. The corrector update can be derived as a regularized LS criterion, which will also be useful to account for the sparsity attribute. To show this, view $\hat{\mathbf{x}}_{k|k-1}$ as a noisy measurement of $\mathbf{x}_k$. It follows readily from (8) that $\hat{\mathbf{x}}_{k|k-1} = \mathbf{x}_k + \mathbf{e}_{k|k-1}$, where $\mathbf{e}_{k|k-1}$ has covariance matrix $\mathbf{P}_{k|k-1}$. Stacking $\hat{\mathbf{x}}_{k|k-1}$ and $\mathbf{y}_k$ to form an augmented measurement vector, yields the following linear regression model

$$\begin{bmatrix} \hat{\mathbf{x}}_{k|k-1} \\ \mathbf{y}_k \end{bmatrix} = \begin{bmatrix} \mathbf{I}_G \\ \mathbf{H} \end{bmatrix} \mathbf{x}_k + \begin{bmatrix} \mathbf{e}_{k|k-1} \\ \mathbf{v}_k \end{bmatrix}$$

where the augmented noise vector has block diagonal covariance matrix denoted as $\mathrm{diag}(\mathbf{P}_{k|k-1}, \mathbf{R}_k)$. The weighted (W)LS estimator for this linear regression problem is given by

$$\hat{\mathbf{x}}_{k|k} = \arg \min_{\mathbf{x}_k \geq \mathbf{0}} \|\hat{\mathbf{x}}_{k|k-1} - \mathbf{x}_k\|_{\mathbf{P}_{k|k-1}^{-1}}^2 + \|\mathbf{y}_k - \mathbf{H}\mathbf{x}_k\|_{\mathbf{R}_k^{-1}}^2 \tag{9}$$

where $\|\mathbf{x}\|_{\mathbf{A}}^2 := \mathbf{x}^T \mathbf{A} \mathbf{x}$. In the absence of non-negativity constraints, the optimal state corrector $\hat{\mathbf{x}}_{k|k}$ can be found in closed form as the cost is quadratic, and likewise its error covariance can be updated as

$$\mathbf{P}_{k|k} = \mathbf{P}_{k|k-1} - \mathbf{P}_{k|k-1} \mathbf{H}^T (\mathbf{H} \mathbf{P}_{k|k-1} \mathbf{H}^T + \mathbf{R}_k)^{-1} \mathbf{H} \mathbf{P}_{k|k-1}. \tag{10}$$

A gradient projection algorithm will be developed in Section IV to solve (9) under non-negativity constraints on the state vector. However, (10) will still be used bearing in mind that this update is approximate now. The TSSG–KF tracker implemented by (8)-(10) is sparsity-agnostic, as it does not explicitly utilize the prior knowledge that $\mathbf{x}_k$ is sparse.

## IV. Sparsity-aware KF Trackers

Taking into account sparsity, this section develops sparsity-cognizant trackers. To this end, the degree of sparsity quantified by the number of nonzero entries of $\mathbf{x}_k$, namely the $\ell_0$-norm $\|\mathbf{x}_k\|_0$, can be used to regularize the LS cost of the previous section. Unfortunately, similar to compressed sensing formulations for solving under-determined linear systems of equations [10], such a regularization results in a non-convex optimization problem that is NP-hard to solve, and motivates relaxing the $\ell_0$-norm with its closest convex approximation, namely the $\ell_1$-norm. Thus, the proposed sparsity-cognizant tracker is based on the state corrector minimizing the following $\ell_1$-regularized WLS cost function

$$\hat{\mathbf{x}}_{k|k} = \arg \min_{\mathbf{x}_k \geq \mathbf{0}} J(\mathbf{x}_k) \tag{11}$$

$$J(\mathbf{x}_k) := \|\hat{\mathbf{x}}_{k|k-1} - \mathbf{x}_k\|^2_{\mathbf{P}^{-1}_{k|k-1}} + \|\mathbf{y}_k - \mathbf{H}\mathbf{x}_k\|^2_{\mathbf{R}^{-1}_k} + 2\lambda_k\|\mathbf{x}_k\|_1.$$

The state corrector minimizing (11), together with the covariance update[1] in (10) and the prediction step in (8), form the recursions of the sparsity-aware TSSG–KF tracker. Relevant design choices and algorithms for minimizing (11) will be elaborated in the next subsection.

The TSSG-KF trackers in (9) and (11) involve both prediction and correction steps, which interestingly can be combined into a single estimation step. Considering that both $\mathbf{x}_{k-1}$ and $\mathbf{x}_k$ are sparse and non-negative, and combining the LS terms for both the prediction and correction steps, the following optimization problem arises for some non-negative $\lambda_{k-1}$ and $\lambda_k$ parameters:

$$\hat{\mathbf{x}}_{k|k} = \arg \min_{\mathbf{x}_{k-1}, \mathbf{x}_k \geq \mathbf{0}} \left\{ \|\hat{\mathbf{x}}_{k-1|k-1} - \mathbf{x}_{k-1}\|^2_{\mathbf{P}^{-1}_{k-1|k-1}} + \|\mathbf{x}_k - \mathbf{F}_k\mathbf{x}_{k-1}\|^2_{\mathbf{Q}^{-1}_k} \right.$$
$$\left. + \|\mathbf{y}_k - \mathbf{H}\mathbf{x}_k\|^2_{\mathbf{R}^{-1}_k} + \lambda_{k-1}\|\mathbf{x}_{k-1}\|_1 + \lambda_k\|\mathbf{x}_k\|_1 \right\}. \tag{12}$$

The performance gain of this tracker was evaluated via simulations and no substantial improvement over the TSSG-KF tracker was observed. For this reason, focus henceforth will be placed on the TSSG-KF tracker in (11).

## A. Parameter selection

The scalar parameter $\lambda_k$ in (11) controls the sparsity-bias tradeoff [17]. The corrector $\hat{\mathbf{x}}_{k|k}$ becomes increasingly sparse as $\lambda_k$ increases, and eventually vanishes, i.e., $\hat{\mathbf{x}}_{k|k} = \mathbf{0}$, when $\lambda_k$ exceeds an upper bound $\bar{\lambda}_k$. There are two systematic means of selecting $\lambda_k$. The first one popular for variable selection in linear regressions is cross-validation [17, pp. 241-249]. The second one is the so-termed absolute variance deviation based selection that has been advocated in the context of outlier rejection setups [16]. Both approaches require solving (11) for different trial values of $\lambda_k$. Even though warm starts reduce the computational burden considerably, this can be certainly affordable for offline solvers of a linear regression problem or a fixed-interval smoothing scenario, but may incur prohibitive delays for real-time applications. For the tracking problem at hand, the simple rule advocated is to set $\lambda_k = \alpha\bar{\lambda}_k$, where $\alpha \in (0, 1)$ is a fixed scaling value to avoid the trivial solution $\hat{\mathbf{x}}_{k|k} = \mathbf{0}$. The bound $\bar{\lambda}_k$ is derived next.

---

[1]A more accurate covariance update will be derived in (28).

**Proposition 1.** *The solution to* (11) *reduces to* $\hat{\mathbf{x}}_{k|k} = \mathbf{0}$ *for any scalar* $\lambda_k \geq \bar{\lambda}_k$, *where*

$$\bar{\lambda}_k = \|\mathbf{P}_{k|k-1}^{-1}\hat{\mathbf{x}}_{k|k-1} + \mathbf{H}^T\mathbf{R}_k^{-1}\mathbf{y}_k\|_\infty. \tag{13}$$

*Proof:* Since $\mathbf{x}_k \geq \mathbf{0}$, it holds that $\|\mathbf{x}_k\|_1 = \mathbf{x}_k^T\mathbf{1}$, where $\mathbf{1}$ denotes the all-one vector. Therefore, $J(\mathbf{x})$ in (11) is differentiable and results in a convex problem. The necessary and sufficient optimality condition states that $\mathbf{x}^*$ is an optimum point iff $(\mathbf{y} - \mathbf{x}^*)^T\nabla J(\mathbf{x}^*) \geq 0$, $\forall \mathbf{y} \geq \mathbf{0}$. For $\mathbf{x}^* = \mathbf{0}$, this condition holds iff $\nabla J(\mathbf{x}^*) \geq \mathbf{0}$. It then follows from (11) that

$$\nabla J(\mathbf{x}) = 2\Big(-\mathbf{P}_{k|k-1}^{-1}(\hat{\mathbf{x}}_{k|k-1} - \mathbf{x}) - \mathbf{H}^T\mathbf{R}_k^{-1}(\mathbf{y}_k - \mathbf{H}\mathbf{x}) + \lambda_k\mathbf{1}\Big). \tag{14}$$

Therefore, $\mathbf{x}^* = \mathbf{0}$ is an optimal solution iff (13) holds. $\qquad\qquad\square$

### B. Gradient projection algorithms

As (11) is a convex problem, convex optimization software such as SeDuMi [26] can be utilized to solve it efficiently. In addition to these solvers, low-complexity iterative methods are developed here, by adopting the gradient projection (GP) algorithms in [7, pp. 212-217]. Note that the proposed algorithms can be used to obtain the sparsity-agnostic tracker from (9) too, since the latter is obtained by minimizing a special case of (11) corresponding to $\lambda_k = 0$.

At each time $k$, the GP is initialized with $\hat{\mathbf{x}}_{k|k}(0) = \hat{\mathbf{x}}_{k|k-1}$ at iteration $l = 0$. The state corrector iterates from $l$ to $(l+1)$ as follows

$$\hat{\mathbf{x}}_{k|k}(l+1) = \big[\hat{\mathbf{x}}_{k|k}(l) - \gamma\nabla J\big(\hat{\mathbf{x}}_{k|k}(l)\big)\big]^+ \tag{15}$$

where $[x]^+$ denotes the projection onto the non-negative orthant, $\gamma$ is the step size, and $\nabla J$ is as in (14). Here $J(\mathbf{x}_k)$ is differentiable because $\|\mathbf{x}_k\|_1 = \mathbf{x}_k^T\mathbf{1}$ when $\mathbf{x}_k \geq \mathbf{0}$.

While (15) amounts to a Jacobi-type iteration updating all the entries at once, one can also devise Gauss-Seidel variants, where entries are updated one at a time [7, pp. 218-219]. This is possible because the non-negative orthant is a constraint set expressible as the Cartesian product of one-dimensional sets, allowing entry-wise updates per iteration $(l+1)$ as

$$\hat{x}_{k|k}^{(j)}(l+1) = \max\Big\{0, \ \hat{x}_{k|k}^{(j)}(l) - \gamma\nabla_j J\big(\tilde{\mathbf{x}}_{k|k}^{(j)}(l)\big)\Big\} \tag{16}$$

where $\tilde{\mathbf{x}}_{k|k}^{(j)}(l) := \big\{\hat{\mathbf{x}}_{k|k}^{(1:j-1)}(l+1), \hat{\mathbf{x}}_{k|k}^{(j:G)}(l)\big\}$ has its first $(j-1)$ entries already updated in the $(l+1)$st iteration. Convergence of the iterations in (16) to the optimum solution of (11) is guaranteed under mild conditions by the results in [7, p. 219]. Specifically, $J(\mathbf{x}_k)$ should be non-negative and its gradient should be Lipschitz continuous, both of which hold for the objective in (11).

**Proposition 2.** *Any limit point of the sequence generated by* (16)*, with arbitrary initialization* $\hat{\mathbf{x}}_{k|k}^{(0)}$*, is an optimal solution of* (11) *provided that the step size* $\gamma$ *is chosen small enough.*

In practice, only a few gradient-projection iterations are run per time step $k$ to allow for real-time sparsity-aware KF tracking.

## V. Enhanced Sparsity-aware IEKF Tracking

The proposed sparsity-aware tracker employs the KF covariance recursion in (10) to update the error covariance of the corrector state estimate. As it does not account for the $\ell_1$-norm regularization, this update is approximate. In order to incorporate the prior knowledge of sparsity when updating the corrector covariance, this section develops an EKF-based approach, which leads to enhanced tracking performance.

Toward this objective, the prior information on sparsity is viewed as an extra measurement $\mu_k = \|\mathbf{x}_k\|_0$, rather than as a regularizing term in the LS cost function. When the number of targets $M$ is known, an apparent choice is to set $\mu_k = M$. Accordingly, tracking will be carried out based on an augmented $(N+1) \times 1$ measurement vector, given by

$$\bar{\mathbf{y}}_k := [\mathbf{y}_k^T \ \mu_k]^T.$$

### A. Viewing sparsity as an extra measurement

The added measurement can be modeled in a general form as

$$\mu_k = \rho(\mathbf{x}_k) + u_k$$

where $\rho(\mathbf{x}_k)$ is a differentiable function approximating the sparsity-inducing $\ell_0$-norm, and $u_k$ denotes zero-mean noise with variance $\sigma_k^2$. The noise term captures both the uncertainty in approximating $\|\mathbf{x}_k\|_0$, as well as the error in attaining the desired degree of sparsity. As to $\rho(\mathbf{x}_k)$, three well-known approximants of the $\ell_0$-norm are the $\ell_1$-norm, the logarithm, and the inverse Gaussian functions:

$$(\ell_1\text{-norm}) \qquad \rho(\mathbf{x}_k) = \mathbf{x}_k^T \mathbf{1}$$

$$(\text{logarithm}) \qquad \rho(\mathbf{x}_k) = \sum_{j=1}^G \log\left(x_k^{(j)} + \delta\right)$$

$$(\text{inverse Gaussian}) \qquad \rho(\mathbf{x}_k) = \sum_{j=1}^G \left(1 - \exp\left(-\frac{(x_k^{(j)})^2}{2\sigma_p^2}\right)\right)$$

where $\delta$ and $\sigma_p$ are tuning parameters, and only $\mathbf{x}_k \geq \mathbf{0}$ is considered. These nonlinear functions are plotted along with the $\ell_0$-norm function for a scalar $\mathbf{x}_k$ in Fig. 1. It can be seen that they all have relatively sharp edges around the origin to approximate the $\ell_0$-norm.

Adding the extra measurement $\mu_k$, the state space model in (4) is augmented to

$$\mathbf{x}_k = \mathbf{F}_k \mathbf{x}_{k-1} \tag{17a}$$

$$\bar{\mathbf{y}}_k = \bar{\mathbf{h}}(\mathbf{x}_k) + \bar{\mathbf{v}}_k \tag{17b}$$

where $\bar{\mathbf{h}}(\mathbf{x}_k) := [(\mathbf{H}\mathbf{x}_k)^T, \ \rho_k(\mathbf{x}_k)]^T$ consists of $N+1$ scalar measurement functions that can be nonlinear in general, and $\bar{\mathbf{v}}_k := [\mathbf{v}_k^T, \ u_k]^T$ has covariance $\bar{\mathbf{R}}_k := \text{diag}(\mathbf{R}_k, \sigma_k^2)$. Similar to (11), the model in (17) leads to a nonlinear (N)LS problem

$$\hat{\mathbf{x}}_{k|k} = \arg \min_{\mathbf{x}_k \geq \mathbf{0}} J_1(\mathbf{x}_k) \tag{18}$$

$$J_1(\mathbf{x}_k) := \|\hat{\mathbf{x}}_{k|k-1} - \mathbf{x}_k\|_{\mathbf{P}_{k|k-1}^{-1}}^2 + \|\mathbf{y}_k - \mathbf{H}\mathbf{x}_k\|_{\bar{\mathbf{R}}_k^{-1}}^2 + \sigma_k^{-2} \left(\mu_k - \rho(\mathbf{x}_k)\right)^2 .$$

Compared with (11), (18) replaces the $\ell_1$-norm of $\mathbf{x}_k$ with an alternative LS-error regularization involving the extra measurement which accounts for the sparsity present. Because (18) directly results from (17), the error covariance of state estimates can be updated using the KF-like recursions developed next.

### B. IEKF algorithm for nonlinear measurement models

Since the augmented $\bar{\mathbf{y}}_k$ in (17b) is a nonlinear function of the wanted TSSG state, the EKF approach is adopted here to update the error covariance along the lines of e.g., [4, Chap. 10]. Specifically, an iterated (I)EKF algorithm is employed, which is tantamount to applying Gauss-Newton iterations to a relevant NLS regression problem [6].

The prediction step of the IEKF is similar to KF, hence $\hat{\mathbf{x}}_{k|k-1}$ and $\hat{\mathbf{P}}_{k|k-1}$ follow directly from the state space model in (17), and coincide with (8). For the correction step per time $k$, IEKF recursions are initialized with $\hat{\mathbf{x}}_{k|k}(0) = \hat{\mathbf{x}}_{k|k-1}$ for $l = 0$, and subsequent iterations proceed as follows [28, Appendix C]

$$\hat{\mathbf{x}}_{k|k}(l+1) = \hat{\mathbf{x}}_{k|k-1} + \mathbf{K}(l)\left(\bar{\mathbf{y}}_k - \bar{\mathbf{h}}(\hat{\mathbf{x}}_{k|k-1}) + \mathbf{\Phi}(l)(\hat{\mathbf{x}}_{k|k}(l) - \hat{\mathbf{x}}_{k|k-1})\right)$$

$$\mathbf{K}(l) = \mathbf{P}_{k|k-1}\mathbf{\Phi}^T(l)\left(\mathbf{\Phi}(l)\mathbf{P}_{k|k-1}\mathbf{\Phi}^T(l) + \bar{\mathbf{R}}_k\right)^{-1} \tag{19}$$

where $\mathbf{\Phi}(l) := \nabla \bar{\mathbf{h}}\big(\hat{\mathbf{x}}_{k|k}(l)\big)^T$ denotes the Jacobian matrix of $\bar{\mathbf{h}}(\cdot)$ evaluated at $\hat{\mathbf{x}}_{k|k}(l)$. After the IEKF iterations are completed at $l = L$, the corrector's error covariance matrix is updated as

$$\mathbf{P}_{k|k} = \mathbf{P}_{k|k-1} - \mathbf{K}(L)\mathbf{\Phi}(L)\mathbf{P}_{k|k-1}. \tag{20}$$

The ensuing proposition establishes the link between IEKF and Gauss-Newton iterations for the related NLS problem.

**Proposition 3.** *Consider the NLS problem [cf. (17) and (18)]*

$$\hat{\mathbf{x}}_{k|k} = \arg \min_{\mathbf{x}_k} \|\hat{\mathbf{x}}_{k|k-1} - \mathbf{x}_k\|^2_{\mathbf{P}^{-1}_{k|k-1}} + \|\bar{\mathbf{y}}_k - \bar{\mathbf{h}}(\mathbf{x}_k)\|^2_{\bar{\mathbf{R}}^{-1}_k}. \tag{21}$$

*Solving* (21) *via Gauss-Newton iterations initialized with* $\hat{\mathbf{x}}_{k|k}(0) = \hat{\mathbf{x}}_{k|k-1}$, *amounts to the IEKF recursions in* (19).

*Proof:* The quadratic terms in (21) can be rewritten as

$$\hat{\mathbf{x}}_{k|k} = \arg \min_{\mathbf{x}_k} \|\mathbf{g}(\mathbf{x}_k)\|^2_2 \tag{22}$$

$$\text{where} \quad \mathbf{g}(\mathbf{x}_k) = \begin{bmatrix} \mathbf{P}^{-1/2}_{k|k-1}(\hat{\mathbf{x}}_{k|k-1} - \mathbf{x}_k) \\ \\ \bar{\mathbf{R}}^{-1/2}_k(\bar{\mathbf{y}}_k - \bar{\mathbf{h}}(\mathbf{x}_k)) \end{bmatrix}. \tag{23}$$

Gauss-Newton iterations for (22) become

$$\hat{\mathbf{x}}_{k|k}(l+1) = \hat{\mathbf{x}}_{k|k}(l) - \left(\mathbf{\Psi}(l)\mathbf{\Psi}^T(l)\right)^{-1} \mathbf{\Psi}(l)\mathbf{g}(\hat{\mathbf{x}}_{k|k}(l)) \tag{24}$$

where $\mathbf{\Psi}(l) := \nabla \mathbf{g}(\hat{\mathbf{x}}_{k|k}(l))$ is the Jacobian transpose evaluated at $\hat{\mathbf{x}}_{k|k}(l)$. Substituting $\mathbf{g}(.)$ from (23) into (24), and applying the matrix inversion lemma to invert the matrix in (24), yields (19) after straightforward algebraic manipulations. $\square$

When Gauss-Newton iterations in (24) are adopted in lieu of IEKF, the resulting error covariance matrix is a function of $\nabla \mathbf{g}$ at the last iteration $L$ given by

$$\mathbf{P}_{k|k} = \left(\mathbf{\Psi}(L)\mathbf{\Psi}^T(L)\right)^{-1}. \tag{25}$$

The sparsity-aware EKF formulation in (18) is a special case of the general NLS problem in (21) corresponding to $\bar{\mathbf{h}}(\mathbf{x}_k) := [(\mathbf{H}\mathbf{x}_k)^T, \ \rho_k(\mathbf{x}_k)]^T$. As a result, the error covariance for the state estimate of (18) can be derived from (25) as

$$\mathbf{P}_{k|k} = \left(\mathbf{P}^{-1}_{k|k-1} + \mathbf{H}^T \mathbf{R}^{-1}_k \mathbf{H} + \frac{1}{\sigma^2_k}\nabla \rho(\hat{\mathbf{x}}_{k|k}(L))\nabla \rho(\hat{\mathbf{x}}_{k|k}(L))^T\right)^{-1}. \tag{26}$$

Compared with (10) for the sparsity-agnostic KF, the last summand in (26) captures the effect of the sparsity-promoting penalty term on the error covariance.

To enforce the non-negativity constraints in (18), one can project each Gauss-Newton iterate in (24) onto the non-negative orthant. Unfortunately, this may not generate a convergent sequence [7, p. 215]. To ensure convergence, the projection should be with respect to a different distance metric than the usual Euclidean distance. Upon defining $\mathbf{B}(l) := (\mathbf{\Psi}(l)\mathbf{\Psi}^T(l))^{-1}$, one implements

$$\hat{\mathbf{x}}_{k|k}(l+1) = \left[\hat{\mathbf{x}}_{k|k}(l) - \left(\mathbf{\Psi}(l)\mathbf{\Psi}^T(l)\right)^{-1}\mathbf{\Psi}(l)\mathbf{g}(\hat{\mathbf{x}}_{k|k}(l))\right]^+_{\mathbf{B}(l)} \tag{27}$$

where $[.]_{\mathbf{B}}^+$ denotes projection onto the non-negative orthant, which minimizes the $\|.\|_{\mathbf{B}}^2$ distance instead of the usual $\|.\|_2^2$. If $\rho(\mathbf{x}_k) = \mathbf{x}_k^T \mathbf{1}$, which is equivalent to the $\ell_1$-norm for $\mathbf{x}_k \geq \mathbf{0}$, then (18) becomes convex and general-purpose convex solvers such as SeDuMi can also be utilized to solve it [26].

The iterative updates in (27) and (26), along with the prediction step (8), constitute the *sparsity-aware TSSG–IEKF tracker*.

### C. Enhanced sparsity-aware KF tracker

As a final note, the sparsity-aware TSSG–KF tracker in Section IV can be enhanced by also casting the $\ell_1$-regularized WLS cost in (11) as an NLS cost. The $\ell_1$-norm term in (11) can be equivalently expressed as an extra LS error term for the extra measurement $0 = \sqrt{2\lambda}\sqrt{\mathbf{x}_k^T \mathbf{1}} + u_k$, where $u_k$ is zero-mean noise with unit-variance. The corresponding covariance update can be derived from (25) as

$$\mathbf{P}_{k|k} = \left( \mathbf{P}_{k|k-1}^{-1} + \mathbf{H}^T \mathbf{R}_k^{-1} \mathbf{H} + \frac{\lambda}{2\mathbf{x}_k^T \mathbf{1}} \mathbf{1}\mathbf{1}^T \right)^{-1}. \tag{28}$$

In all, the state update in (11), together with the prediction step in (8) and the refined covariance update in (28), form the recursions of the *enhanced sparsity-aware TSSG–KF tracker*.

## VI. POSITION ESTIMATION AND TRACK FORMATION

The TSSG filters developed so far produce a dynamic TSS map of the operational environment. Such information is adequate to describe the targets' distribution and spatial occupancy over the sensing field of interest, similar in spirit to the PHD filter which portrays the targets' intensity function and the BOF that depicts their occupancy map. In many tracking applications however, more informative estimates such as target positions and trajectories are desired. This section provides TSSG-based solutions to these estimation tasks too.

For the PHD approach, methods performing these extra steps have been reported using particle PHD filters [12], [20], [25], or Gaussian mixture (GM)-PHD filters [24]. Target positions are typically identified by peak-picking the target intensity function being tracked, and the estimated target positions are treated as measurements for the ensuing data association and track recovery tasks. PHD filters view each particle or each Gaussian component involved as a target [22], [30], and employ conventional target movement models to describe the state transition. As a result, most of the well-known data association methods can be run after PHD filtering [3], [9, Chapters 6-7]. Examples include the auction algorithm proposed in [20], and the joint probabilistic data association (JPDAF) algorithm [23]. Likewise for the BOF, the target movement model is employed in updating the HMM filter as well, which makes it feasible to be combined with a well-established data association method such as the JPDAF [23].

In contrast, the TSSG state equation only models the dynamic behavior of the TSS distribution on the grid, in which grid points are not treated as targets, and hence do not directly obey the conventional target movement model. As remarked in Section II, only partial information about position changes is explicitly captured by the state transition matrix $\mathbf{F}_k$, while other factors such as velocity are implicit. Due to this major difference, conventional data association methods cannot be directly adopted as a follow-up to TSSG filtering. This section develops estimators of target positions and tracks for multi-target scenarios, based solely on the limited information regarding target transition probabilities on the grid.

### A. Target position estimation

Given the output $\hat{\mathbf{x}}_{k|k}$ of the TSSG filter, target positions can be obtained from (6) provided that the subset of grid points associated with each target is known in the form of $\mathcal{G}_k^{(m)}$, $\forall m$.

Starting from $\hat{\mathbf{x}}_{k|k}$, one can apply appropriate clustering techniques to identify $\mathcal{G}_k^{(m)}$. When the number of targets $M$ is known, simple parametric clustering methods such as the $k$-means can be used [8, pp. 424–429]. When $M$ is unknown, one can perform joint clustering and model order selection. Such algorithms utilize some global model order selection criteria such as Akaike's information criterion to determine the best number of clusters $\hat{M}$, as well as the clusters $\{\hat{\mathcal{G}}_k^{(m)}\}_{m=1}^{\hat{M}}$ themselves [31]. Other nonparametric clustering methods can be employed as well, without assuming or estimating the number of clusters. For example, hierarchical clustering techniques either aggregate or divide the data based on some proximity measure, while density estimation-based nonparametric approaches identify clusters and their number from the modes of the empirical density function of the unknowns; see e.g., [18] for a survey.

Having acquired $\hat{M}$ and $\{\hat{\mathcal{G}}_k^{(m)}\}_{m=1}^{\hat{M}}$, and based on (6), the target positions can be obtained individually from the TSSG estimates on the associated clusters of grid points $\forall i \in \hat{\mathcal{G}}_k^{(m)}$, as follows:

$$\hat{\mathbf{p}}_k^{(m)} = \frac{\sum_{i \in \hat{\mathcal{G}}_k^{(m)}} \mathbf{g}_i \hat{x}_{k|k}^{(i)}}{\sum_{i \in \hat{\mathcal{G}}_k^{(m)}} \hat{x}_{k|k}^{(i)}}, \quad m = 1, 2, \ldots, \hat{M}. \tag{29}$$

### B. Position-to-track association

Suppose that there are $M_t$ tracks from time slot 1 up to $k-1$, and $\hat{\mathbf{p}}_{k-1}^{(m)}$ has been associated with track $t$ and hence alternatively expressed as $\hat{\mathbf{p}}_{k-1}^{(t)}$, $t = 1, \ldots, M_t$. The goal of track association is to assign the position estimates $\left\{\hat{\mathbf{p}}_k^{(m)}\right\}_{m=1}^{M}$ of the $M$ targets at time $k$ to one of the established $M_t$ tracks. For clarity in exposition, suppose first that $M = M_t$ and there is no target birth or death. This assumption will be removed later on. Evidently, there are $M!$ different assignments, which must be examined to find the best possible association.

Given $\mathbf{y}_{1:k-1}$, the first step is to establish a track prediction model to be used for computing the predicted track positions $\{\hat{\mathbf{p}}_{k|k-1}^{(t)}\}_{t=1}^{M_t}$ and their error covariances. Note from (29) that the target position estimates conditioned on the TSSG are independent of the per-sensor measurements. Hence, it suffices to predict $\{\hat{\mathbf{p}}_{k|k-1}^{(t)}\}_t$ solely from the TSSG vector $\hat{\mathbf{x}}_{k-1|k-1}$. To do so, focus on track $t$ and form a $G \times 1$ vector $\breve{\mathbf{x}}_{k-1,t}$ that only retains the entries of $\hat{\mathbf{x}}_{k-1|k-1}$ belonging to the $t$-th cluster of grid points in $\mathcal{G}_{k-1}^{(t)}$; that is, $\breve{x}_{k-1,t}^{(j)} = \hat{x}_{k-1}^{(j)}$ for $j \in \mathcal{G}_{k-1}^{(t)}$ and $\breve{x}_{k-1,t}^{(j)} = 0$ otherwise, $\forall j$.

Given $\breve{\mathbf{x}}_{k-1,t}$ at time $k-1$, the predicted TSSG belonging to track $t$ at time $k$ becomes

$$\breve{\mathbf{x}}_{k|k-1,t} = \mathbf{F}_k \breve{\mathbf{x}}_{k-1,t}$$

and correspondingly, the predicted track position is

$$\hat{\mathbf{p}}_{k|k-1}^{(t)} = \frac{\sum_{j=1}^{G} \mathbf{g}_j \breve{x}_{k|k-1,t}^{(j)}}{\sum_{j=1}^{G} \breve{x}_{k|k-1,t}^{(j)}} \ . \tag{30}$$

The normalized quantities $\breve{x}_{k|k-1,t}^{(j)}/(\sum_{j=1}^{G} \breve{x}_{k|k-1,t}^{(j)})$ in (30) play the role of fractional weights when the corresponding grid positions $\mathbf{g}_j$ are used to estimate the track position. Viewing $\hat{\mathbf{p}}_{k|k-1}^{(t)}$ as the weighted average of $G$ position-samples $\{\mathbf{g}_j\}_{j=1}^{G}$, it is straightforward to estimate the covariance of $\hat{\mathbf{p}}_{k|k-1}^{(t)}$ using the sample covariance, as

$$\hat{\mathbf{P}}_{k|k-1}^{(t)} = \frac{\sum_{j=1}^{G} \breve{x}_{k|k-1,t}^{(j)} (\mathbf{g}_j - \hat{\mathbf{p}}_{k|k-1}^{(t)})(\mathbf{g}_j - \hat{\mathbf{p}}_{k|k-1}^{(t)})^T}{\sum_{j=1}^{G} \breve{x}_{k|k-1,t}^{(j)}} \ . \tag{31}$$

The process in (30)-(31) is repeated for all target tracks $t = 1, \ldots, M_t$, so that the prediction estimates and covariances become available for all tracks.

Now, the aim is to associate the predicted track positions $\{\hat{\mathbf{p}}_{k|k-1}^{(t)}\}_t$ in (30) with the target position estimates $\{\hat{\mathbf{p}}_{k}^{(t)}\}_m$ in (29). To this end, define the decision variables $a(t,m) \in \{0,1\}$ for $t = 1, \ldots, M_t$ and $m = 1, \ldots, M$, where $a(t,m) = 1$ amounts to deciding that target $m$ measured at $\hat{\mathbf{p}}_k^{(m)}$ is assigned to track $t$. The pairwise-association cost can be quantified using the Mahalanobis distance between track $t$'s prediction and $\hat{\mathbf{p}}_k^{(m)}$ as a measurement, that is

$$\mathrm{MD}(t,m) := \left(\hat{\mathbf{p}}_{k|k-1}^{(t)} - \hat{\mathbf{p}}_k^{(m)}\right)^T (\hat{\mathbf{P}}_{k|k-1}^{(t)})^{-1} \left(\hat{\mathbf{p}}_{k|k-1}^{(t)} - \hat{\mathbf{p}}_k^{(m)}\right) \ . \tag{32}$$

The following optimization problem is formulated to minimize the total association cost subject to linear constraints that ensure one-to-one track-to-measurement mapping:

$$\min_{a(t,m)\in\{0,1\}} \quad \sum_{t=1}^{M}\sum_{m=1}^{M} a(t,m)\mathrm{MD}(t,m) \tag{33}$$

$$\text{such that} \quad \sum_{m=1}^{M} a(t,m) = 1, \ \forall t = 1,\dots,M_t,$$

$$\sum_{t=1}^{M} a(t,m) = 1, \ \forall m = 1,\dots,M.$$

It is worth mentioning that (33) is a special case of the so called assignment problem, which is a well-known data association algorithm [9, pp. 342–349]. Its solution can be efficiently computed in polynomial time using integer programming solvers such as the Hungarian algorithm [19].

The track association problem in (33) can be modified to handle track birth and death scenaria [9]. Toward this objective, introduce a dummy target $m = 0$ and a dummy track $t = 0$. The one-to-one constraints in (33) are modified as follows: each track is assigned to at most one target position measurement, but the dummy track can be associated with any number of targets; meanwhile, each position measurement is assigned to at most one track, but the dummy measurement can be assigned to multiple tracks; further, the dummy target cannot be associated with the dummy track. Such a modified association problem resembles the auction algorithm [9], [20], along with the corresponding association costs defined in (32). The computational burden of this combinatorial problem can be reduced by removing some unlikely association pairs in advance. Essentially, if for a track $t$ all the association costs $\{\mathrm{MD}(t,m)\}_m$ exceed a large threshold, then this track is considered "dead", and is associated with the dummy target. Similarly for a target $m$, if all the association costs $\{\mathrm{MD}(t,m)\}_t$ are too large, then this target is considered "born", and is associated with the dummy track.

Once the position-to-track association is completed, velocity estimates can be obtained too. This is possible by subtracting target position at time $k - 1$ from its position at time $k$ and dividing by the sampling period.

Finally, it is worth noting that in formulating (33), only the state transition probability matrix $\mathbf{F}_k$ is needed, regardless of the underlying target movement model. It is possible however to utilize each target's movement model to develop other (more effective) data association schemes, and refine the track estimates as well. Such association and track refinement steps will take place after every TSSG update, using the output of the TSSG tracker to form the position-measurements (29) for the ensuing parallel target trackers, one for each target. The results will not be fed back to the TSSG trackers, thus ensuring resilience of TSSG estimates to data mis-association and track estimation errors.

## VII. NUMERICAL TESTS

Consider a $300 \times 300$ square-meter surveillance region along with a $10 \times 10$ rectangular grid with equally-spaced grid points. Therefore, each grid cell is of size $30 \times 30$. Simulations are performed for both single- and multi-target scenarios.

### A. Single-target case

A single target starts at the south-west corner of the grid at time $k = 1$, and moves northeast according to a constant velocity model

$$\mathbf{p}_k = \mathbf{p}_{k-1} + \bar{\mathbf{v}} T_s + \mathbf{n}_k$$

where $\bar{\mathbf{v}}$ denotes the target's constant velocity assumed known and given by $\bar{\mathbf{v}} = (15, 15)$ meters per second; $\mathbf{p}_{k-1}$ is the previous target position; $T_s = 1$ is the sampling time in seconds; and $\mathbf{n}_k$ represents modeling noise of zero-mean and variance $\sigma_n^2 \mathbf{I}_2$. Given this model and ignoring $\mathbf{n}_k$, if the target starts at the center of the grid cell it is currently in, then at the next time instant, it will arrive at the northeast corner of this grid cell conjoining the north, east, and northeast grid cells. Due to the symmetrically distributed noise, the target will have equal probability of falling inside each of the 4 grid cells. It is assumed that $\sigma_n$ is small enough so that the probability of a target moving into grid cells other than its four adjacent ones is negligible. The resultant movement model is as follows: a target stays on the current grid point with probability $1/4$, and moves north, east, or northeast with probability $1/4$. Whenever the target moves outside the boundaries of the surveillance region, tracking stops. One random realization of this movement model is plotted in Fig. 2 and is considered for the ensuing simulations starting with the single-target case. The target's signal strength is $s = 10$, and there are $N = 20$ sensors distributed randomly over the surveillance region measuring the received TSS. The measurement noise $\mathbf{v}_k$ is zero-mean Gaussian white with unit variance. The propagation function $h(x)$ in (1) is given by $h(x) = c/(c + x^2)$ for $x \geq 0$, where $c$ is chosen so that $h(60) = 0.5$. Apparently, $h(0) = 1$ and $h(x)$ is monotonically decreasing as $x$ increases.

The proposed sparsity-agnostic and sparsity-aware TSSG-KF trackers in Sections III-IV are employed to estimate the target signal strengths and position vectors over time. The position estimation accuracy is measured by the average root mean-square error (RMSE) in the form of RMSE $= \sqrt{\frac{1}{K_{\max}} \sum_{k=1}^{K_{\max}} \|\hat{\mathbf{p}}_k - \mathbf{p}_k\|_2^2}$, where $K_{\max}$ is the tracking duration and $\hat{\mathbf{p}}_k$ is obtained as in (6). The covariance matrix of the process noise $\mathbf{w}_k$ is set to $\mathbf{Q}_k = \mathbf{I}_G$ in (8), and $1,000$ Monte Carlo runs over the random measurement noise are performed to compute the RMSE. To shed light on the role of the $\ell_1$-norm sparsity penalty term in (11), Fig. 3 depicts the RMSE performance with respect to the sparsity-controlling coefficient $\lambda_k$ as a fraction of $\bar{\lambda}_k$ in (13). The sparsity-agnostic tracker corresponds to setting $\lambda = 0$ in (11), and is also plotted for comparison. It is seen that the sparsity-aware KF tracker outperforms the sparsity-agnostic one for a large range of $\lambda_k \neq 0$ values, and $\lambda_k = 0.1\bar{\lambda}_k$ appears to yield the lowest RMSE for this test. The optimal HMM filter exhibits the best performance, but requires accurate knowledge of the target signal strength.

Fig. 4 depicts the RMSE of the sparsity-aware TSSG-IEKF tracker of (18), with $\mu_k = 1$ and for different values of $\sigma_k$. This tracker incorporates sparsity as an extra measurement, and selects the sparsity model $\rho(\mathbf{x}_k)$ as the $\ell_1$-norm function. Evidently, this extra measurement is effective in promoting sparsity, which leads to improved performance relative to the sparsity-agnostic tracker. The noise variance $\sigma_k^2$ of the sparsity measurement in (17b) is a design parameter chosen in accordance with the sensor measurements (here having unit variance). As Fig. 4 indicates, there is an optimal value of $\sigma_k$ that attains the most effective tradeoff between the sensor measurements and the sparsity-induced measurement. As $\sigma_k$ becomes larger, the tracker collects less information from the extra measurement, and eventually becomes sparsity-agnostic when $\sigma_k$ is too large. On the other hand, when $\sigma_k$ is too small, the tracker is predominantly enforcing a sparse solution without considering much the sensor measurements, which also degrades tracking performance.

Both sparsity-aware TSSG trackers, the TSSG-KF tracker with $\lambda_k = 0.1\bar{\lambda}_k$ and the TSSG-IEKF tracker with $\sigma_k = 2$, are compared in Fig. 5 in terms of their RMSE performance versus time. The curves are generated using $1,000$ Monte Carlo runs. These two sparsity-aware trackers exhibit similar performance, both outperforming the sparsity-agnostic tracker. The clairvoyant optimal HMM filter is also tested as the benchmark.

Finally, Fig. 6 demonstrates the dynamic behavior of the sparsity-aware estimator in (11) with $\lambda_k = 0.9\bar{\lambda}_k$. Even though the sparsity-aware TSSG-KF performs worse than sparsity-agnostic TSSG-KF for this value of $\lambda_k$, it is chosen to demonstrate how sparsity affects the tracking process. The estimated TSSG state vectors are depicted over time, with a circle representing a nonzero TSS at the corresponding grid point. The true and estimated tracks are plotted as well. For clarity, only the projection of the target track on the y-direction is depicted. It is seen that the "cloud" of nonzero target signal strengths follows the true track. The estimated target profile is seen to be indeed spatially sparse. The size of the nonzero support indicates the uncertainty in target position estimates, which apparently does not grow over time, even when using a simple grid-induced linear KF tracker to follow the state transition pattern.

### B. Multi-target case

Two targets are respectively located at the south-center and west-center of the grid at time $k = 1$. They start moving according to the same movement model used for the single-target case. Fig. 7 plots one random realization of these target trajectories used for the ensuing multi-target test cases. Adhering to as1), these two trajectories do not overlap on the same grid point at the same time. The target signal strengths are set to be $s^{(1)} = s^{(2)} = 10$. It is assumed that the trackers know the number of targets unless otherwise stated. There are 100 sensors deployed randomly over the surveillance region to measure the total received signal strengths.

First, the position estimation method presented in Subsection VI-A is tested. Fig. 7 depicts the position estimates as circles along with the true target trajectories, for both the sparsity-agnostic TSSG-KF and the sparsity-aware TSSG-KF trackers with $\lambda_k = 0.1\bar{\lambda}_k$. When the $\ell_1$-norm sparsity-promoting regularization term is not present (cf. Fig. 7), position estimates are rather inaccurate and some of them fall far from either of the two targets. In contrast, the sparsity-aware TSSG-KF in Fig. 7 results in quite accurate position estimates. One can clearly associate each position estimate with one of the two targets, and readily visualize target tracks from the position estimates. Before the position estimates are associated with individual targets, a pertinent performance metric quantifying estimation accuracy is the so-called Wasserstein distance (WD) that measures the distance between two finite sets [13]. Let $P_k = \{\mathbf{p}_k^{(m)}\}_m$ denote the finite set of the true target positions at time $k$ and $\hat{P}_k = \{\hat{\mathbf{p}}_k^{(n)}\}_n$ the set of position estimates, respectively. Let $d(.,.)$ stand for the Euclidean $\ell_2$-norm, and $|\cdot|$ for set cardinality. The $L^p$ WD between these two sets is defined as

$$d_p^W(P_k, \hat{P}_k) = \min_{\{C_{mn}\}} \left( \sum_{\mathbf{p}^{(m)} \in P_k} \sum_{\hat{\mathbf{p}}^{(n)} \in \hat{P}_k} C_{mn} d\left(\mathbf{p}^{(m)}, \hat{\mathbf{p}}^{(n)}\right)^p \right)^{1/p}$$

$$\text{subject to} \quad \sum_{m=1}^{|P_k|} C_{mn} = \frac{1}{|\hat{P}_k|}, \ \forall n = 1, \ldots, |\hat{P}_k|$$

$$\sum_{n=1}^{|\hat{P}_k|} C_{mn} = \frac{1}{|P_k|}, \ \forall m = 1, \ldots, |P_k|.$$

Fig. 8 depicts the $L^1$ WD for both sparsity-aware TSSG-KF and TSSG-IEKF trackers, in comparison with the sparsity-agnostic TSSG-KF tracker. The TSSG-IEKF tracker is implemented with $\mu_k = 2$ and $\sigma_k = 2$. The WD is evaluated by averaging over $1,000$ Monte Carlo runs for each tracker. Evidently, both sparsity-aware designs are effective and improve the WD performance.

The track formation algorithm of Subsection VI-B is investigated next for the same target realization. The target tracks formed using the position estimates of a single Monte Carlo run are plotted in Fig. 9, for the sparsity-agnostic TSSG-KF tracker. The estimated track for target 1 is not even plotted because it deviates too much from the true trajectory. The estimated track for target 2 shows some erratic behavior.

As will be discussed shortly, the unsatisfactory performance is not due to the proposed track formation algorithm itself; rather, it is a manifestation of inaccurate clustering that results from badly shaped TSSG estimates to begin with. The accuracy of the TSS map provided by the TSSG filters is essential in ensuring good performance of position estimates and track formation algorithms. Fig. 9 illustrates the track estimates obtained after processing the sparsity-aware TSSG-KF output. It can be seen that both targets are closely tracked. To compare these methods quantitatively, the RMSE curves for the two targets are plotted versus time in Fig. 10, for $1,000$ Monte Carlo runs. It is evident that exploitation of sparsity markedly improves performance of the TSSG filters. In addition, sparsity-aware TSSG-KF seems to outperform the TSSG-IEKF for this specific setting and choice of parameters.

To further illustrate the importance of TSSG estimation for subsequently forming position and track estimates, Fig. 11a depicts two snapshots of the TSSG heat maps after the KF prediction and correction steps at times $k = 2$ and $3$. For the sparsity-agnostic TSSG-KF tracker, the correction heat map at $k = 2$ seems to contain three clusters while there are only two targets. In the correction heat map at $k = 3$, there is a single point in the lower right which is nonzero and far from both targets. This spurious point can have a detrimental effect during the clustering phase as it can greatly shift mean positions of the two clusters. These malign effects do not show up in the TSSG heat maps for the sparsity-aware TSSG-KF in Fig. 11b, where heat maps exhibit two compact clusters in both KF correction steps.

Lastly, simulations for an unknown number of targets are performed on a $15 \times 15$ grid with the true and estimated target tracks plotted in Fig. 12. In this setup, targets 1 and 2 begin their movement at time $k = 1$; at $k = 5$ target 3 is born, and at $k = 10$ target 1 disappears. The sparsity-aware TSSG-KF is utilized in both simulations. Various clustering options are available when the number of clusters is unknown [31]. Here a simple MATLAB routine called "silhouette" is used to determine the best number of natural clusters in the TSS maps. After $k$-means clustering is performed, silhouette returns a value between $-1$ and $1$ for every point that has participated in the clustering phase. The value that silhouette returns measures how well every point is explained by the cluster it belongs to, compared to other clusters. A value close to $1$ is desirable. Therefore, silhouette values averaged over the clustered points offer a good measure of how well clusters explain the points which belong to them. The number of clusters with the largest average silhouette value is selected as the most appropriate number of clusters. It can be seen that the three targets are accurately tracked. However, a small erroneous track emerges close to target 1 for two time periods. Unfortunately, performance of the case with unknown number of targets is not always as accurate as shown here and more than one inaccurate track may arise. On the other hand, when applied to the two-target example previously considered in the absence of target births or deaths, the algorithm with unknown number of targets is always successful in recovering accurate target tracks.

## VIII. Conclusions

The problem of tracking multiple targets on a plane using the superposition of their received signal strengths as measurements has been investigated. A grid-based state space model was introduced to describe the dynamic behavior of target signal strengths. This model not only renders the nonlinear estimation problem linear, but also facilitates incorporation and exploitation of the grid-induced sparsity present. Two sparsity-aware Kalman trackers were developed to exploit this sparsity attribute: TSSG-KF promoting sparsity of the state estimates through $\ell_1$-norm minimization, and TSSG-IEKF effecting sparsity by viewing it as an extra measurement. To address the challenge of updating the state estimation error covariances under sparsity constraints, a novel approach based on iterative extended KF and measurement augmentation was also developed to provide tractable and accurate covariance updates. Position estimation and position-to-track association issues were considered as well. The proposed trackers do not require knowing the number of targets or their signal strengths, and considerably reduce complexity when compared to the optimal hidden Markov model filter. They offer improved tracking performance at reduced sensing and computational cost, especially when compared to sparsity-agnostic trackers.

**Acknowledgement.** The first two authors wish to thank Dr. D. Angelosante of ABB, Switzerland, and Prof. Stergios I. Roumeliotis of the CS&E Department at the University of Minnesota for helpful discussions and suggestions on the early stages of this work.

## Appendix

From the total probability argument, it holds that

$$p\left(x_k^{(j)} \neq 0 \,\Big|\, j \in \mathcal{G}_k^{(m)}\right) = \sum_{i=1}^{G} p\left(x_k^{(j)} \neq 0, x_{k-1}^{(i)} \neq 0, i \in \mathcal{G}_{k-1}^{(m)} \,\Big|\, j \in \mathcal{G}_k^{(m)}\right)$$

which leads to the following equality after invoking as2) in Bayes' rule[2]:

$$p\left(x_k^{(j)} \neq 0 \,\Big|\, j \in \mathcal{G}_k^{(m)}\right) = \sum_{i=1}^{G} f_k^{(ji)} p\left(x_{k-1}^{(i)} \neq 0, i \in \mathcal{G}_{k-1}^{(m)}\right) = \sum_{i \in \mathcal{G}_{k-1}^{(m)}} f_k^{(ji)} p\left(x_{k-1}^{(i)} \neq 0 \,\Big|\, i \in \mathcal{G}_{k-1}^{(m)}\right).$$

$$(34)$$

---

[2] It holds trivially for the dummy target $m = 0$ as well, because $p(x_k^{(j)} \neq 0 | j \in \mathcal{G}_k^{(0)}) = 0$ and $p(x_k^{(j)} \neq 0, j \in \mathcal{G}_k^{(0)}) = 0$.

Any grid point $j = 1, \ldots, G$ with a nonzero $x_k^{(j)} \neq 0$ is associated with a single target index $m_k^{(j)} \in [1, M]$ at time $k$, which means $p(x_k^{(j)} \neq 0, j \in \mathcal{G}_k^{(m_k^{(j)})}) \neq 0$ for $m_k^{(j)} \in [1, M]$; and according to as1), $p(x_k^{(j)} \neq 0, j \in \mathcal{G}_k^{(m)}) = 0$, $\forall m \neq m_k^{(j)}$ or $m = m_k^{(j)} = 0$. Invoking $p(x_k^{(j)} \neq 0) = \sum_{m=0}^{M} p(x_k^{(j)} \neq 0, j \in \mathcal{G}_k^{(m)})$, and noting that $p(j \in \mathcal{G}_k^{(m_k^{(j)})}) = 1$, yields

$$p(x_k^{(j)} \neq 0) = p(x_k^{(j)} \neq 0, j \in \mathcal{G}_k^{(m_k^{(j)})}) = p(x_k^{(j)} \neq 0 | j \in \mathcal{G}_k^{(m_k^{(j)})}), \quad \forall j. \tag{35}$$

Similarly for a grid point $i$ at time $(k-1)$, there exists a target index $m_{k-1}^{(i)} \in [0, M]$ such that $p(x_{k-1}^{(i)} \neq 0) = p(x_{k-1}^{(i)} \neq 0, i \in \mathcal{G}_{k-1}^{(m_{k-1}^{(i)})})$, and $p(x_{k-1}^{(i)} \neq 0, i \notin \mathcal{G}_{k-1}^{(m_{k-1}^{(i)})}) = 0$, $\forall i \in [1, G]$. Under as1) and as2), it follows from (34) and (35) that

$$
\begin{aligned}
x_k^{(j)} &= s^{(m_k^{(j)})} p\left(x_k^{(j)} \neq 0\right) = s^{(m_k^{(j)})} p\left(x_k^{(j)} \neq 0 \,\Big|\, j \in \mathcal{G}_k^{(m_k^{(j)})}\right) = s^{(m_k^{(j)})} \sum_{i=1}^{G} f_k^{(ji)} p\left(x_{k-1}^{(i)} \neq 0, i \in \mathcal{G}_{k-1}^{(m_k^{(j)})}\right) \\
&= \sum_{\forall i: m_{k-1}^{(i)} = m_k^{(j)}} f_k^{(ji)} s^{(m_{k-1}^{(i)})} p\left(x_{k-1}^{(i)} \neq 0, i \in \mathcal{G}_{k-1}^{(m_{k-1}^{(i)})}\right) + s^{(m_k^{(j)})} \sum_{\forall i: m_{k-1}^{(i)} \neq m_k^{(j)}} f_k^{(ji)} \underbrace{p\left(x_{k-1}^{(i)} \neq 0, i \notin \mathcal{G}_{k-1}^{(m_{k-1}^{(i)})}\right)}_{=0, \ \forall i} \\
&= \sum_{\forall i: m_{k-1}^{(i)} = m_k^{(j)}} f_k^{(ji)} s^{(m_{k-1}^{(i)})} p\left(x_{k-1}^{(i)} \neq 0\right) + \sum_{\forall i: m_{k-1}^{(i)} \neq m_k^{(j)}, m_{k-1}^{(i)} = 0} f_k^{(ji)} s^{(0)} \underbrace{p\left(x_{k-1}^{(i)} \neq 0, i \in \mathcal{G}_{k-1}^{(m_{k-1}^{(i)})}\right)}_{=0, \ \forall i: \ m_{k-1}^{(i)} = 0} \\
&= \sum_{\forall i: m_{k-1}^{(i)} = m_k^{(j)}} f_k^{(ji)} s^{(m_{k-1}^{(i)})} p\left(x_{k-1}^{(i)} \neq 0\right) + \sum_{\forall i: m_{k-1}^{(i)} \neq m_k^{(j)}, m_{k-1}^{(i)} = 0} f_k^{(ji)} s^{(m_{k-1}^{(i)})} p\left(x_{k-1}^{(i)} \neq 0 | m_{k-1}^{(i)} = 0\right) \\
&= \sum_{i=1}^{G} f_k^{(ji)} x_{k-1}^{(i)}, \quad \forall j \in [1, G].
\end{aligned}
\tag{36}
$$

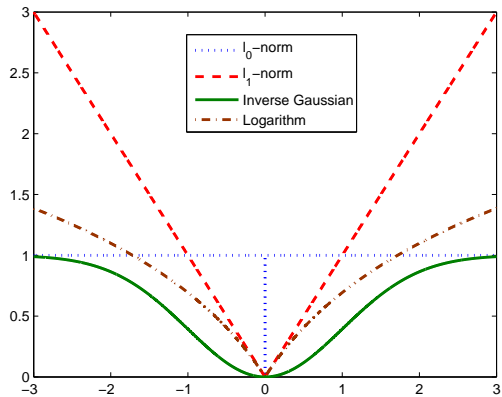

Fig. 1: The $\ell_0$-norm and its three approximations.

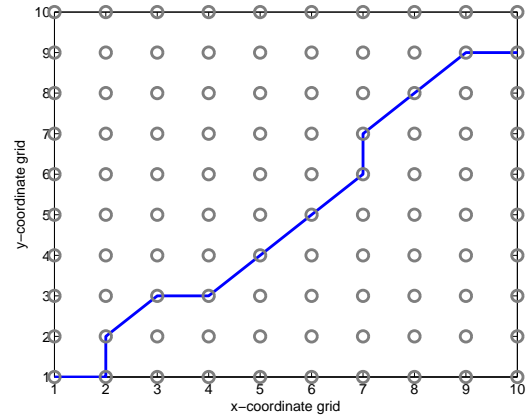

Fig. 2: True target track on the grid.

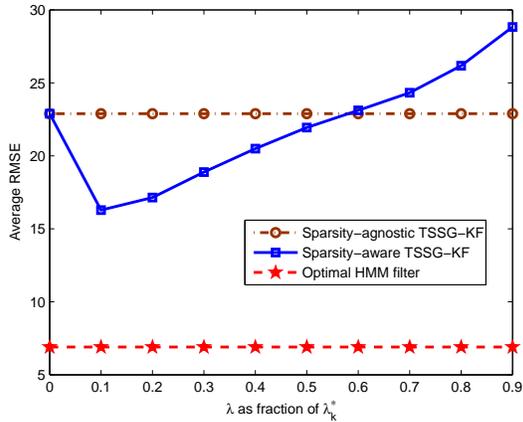

Fig. 3: Sparsity-agnostic and sparsity-aware TSSG-KF trackers.

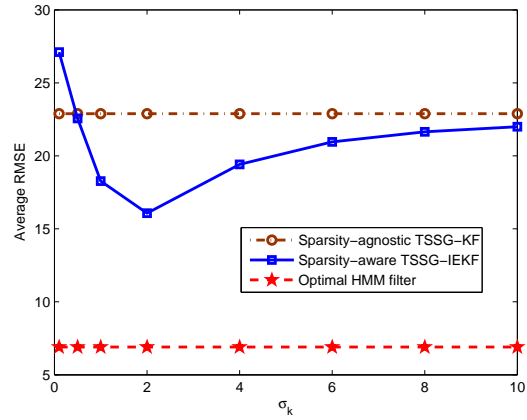

Fig. 4: TSSG-IEKF tracker with an extra sparsity measurement.

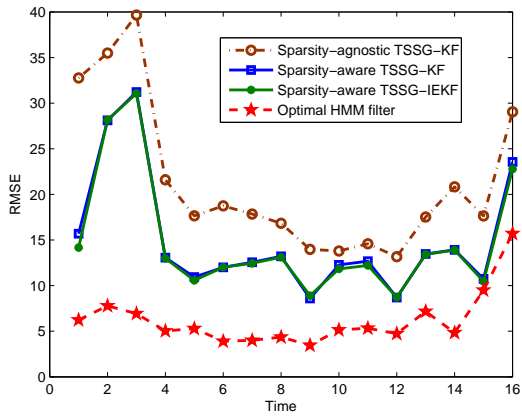

Fig. 5: Comparison of TSSG-KF and TSSG-IEKF trackers.

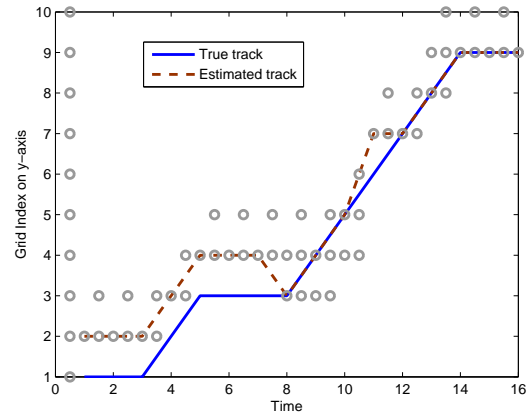

Fig. 6: Nonzero support of estimated TSSG, true, and estimated tracks (y-direction only).

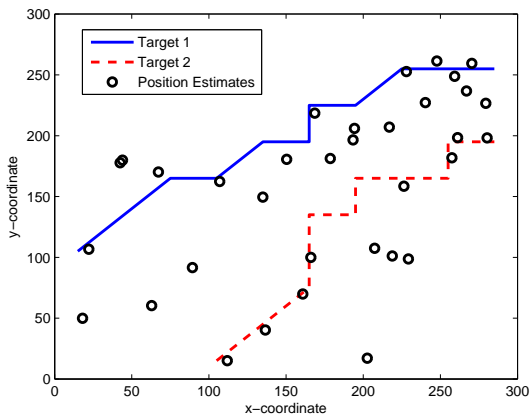

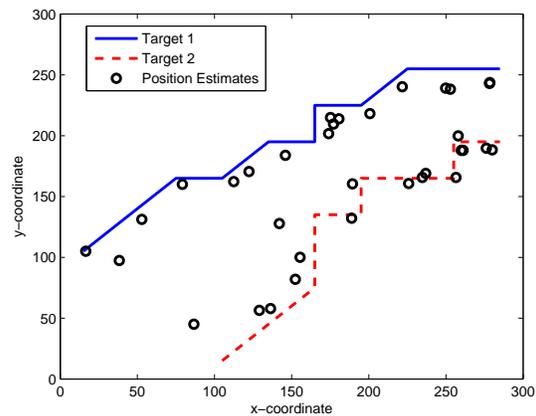

Fig. 7: True tracks and position estimates for two targets: (left) sparsity-agnostic TSSG-KF tracker, (right) sparsity-aware TSSG-KF tracker. Circles indicate the estimated target positions.

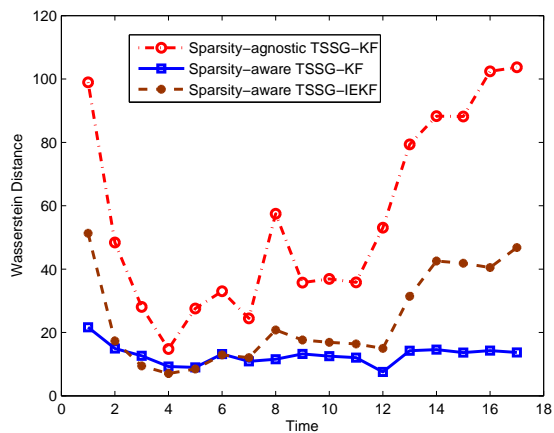

Fig. 8: WD versus time for sparsity-agnostic and sparsity-aware TSSG-KF and TSSG-IEKF trackers.

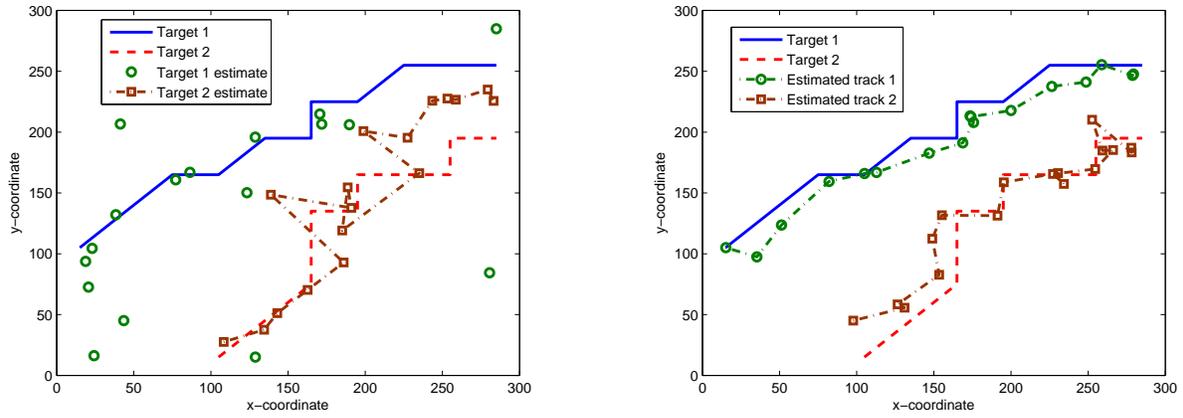

Fig. 9: True and estimated tracks: (left) sparsity-agnostic TSSG-KF; (right) sparsity-aware TSSG-KF;

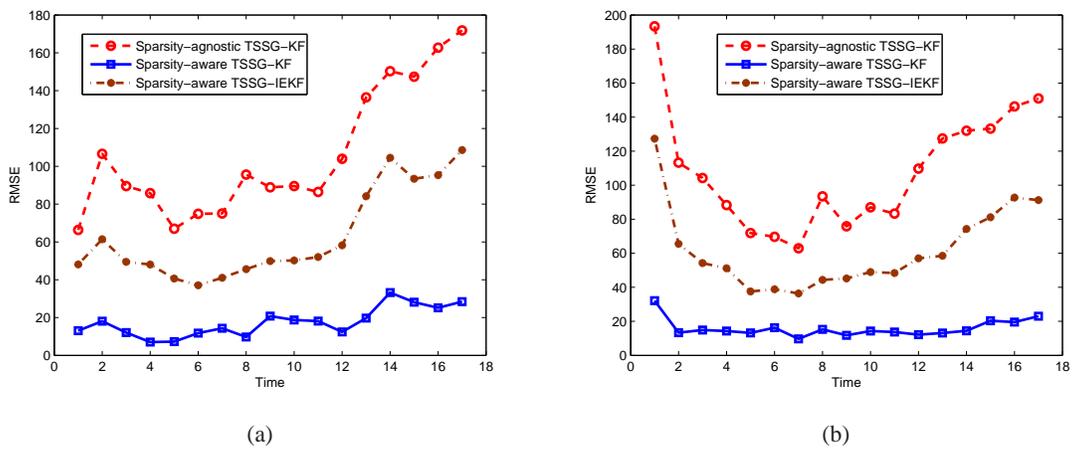

Fig. 10: Tracking performance for multi-target case: (a) RMSE for target 1, (b) RMSE for target 2.

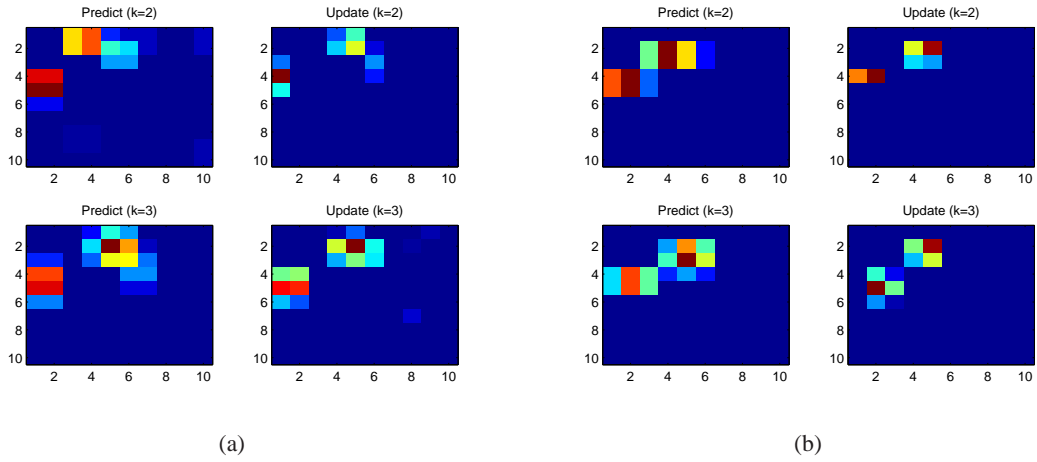

(a)                                                                 (b)

Fig. 11: Heat map: (a) sparsity-agnostic TSSG-KF tracker, (b) sparsity-aware TSSG-KF tracker.

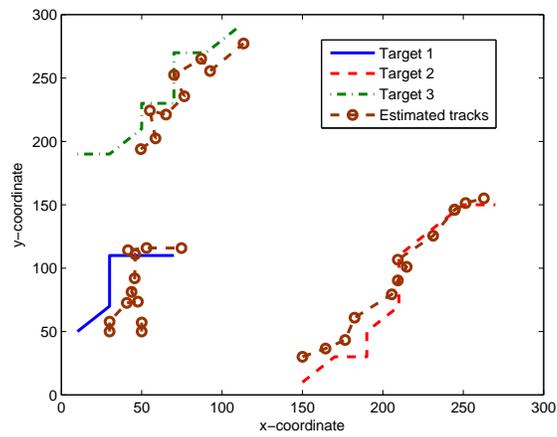

Fig. 12: True and estimated tracks with unknown number of clusters.